\documentstyle[prb,aps,twocolumn]{revtex}
\def\geqap{\,\raise 2pt \hbox{$>\kern-11pt \lower 5pt \hbox{$\sim$}$}\,}
\def\leqap{\,\raise 2pt \hbox{$<\kern-10pt \lower 5pt \hbox{$\sim$}$}\,}
\begin{document}
\draft
\title{Phase Diagram in Manganese Oxides}
\author{Ryo Maezono, Sumio Ishihara$^{*}$, and  Naoto Nagaosa}
\address{Department of Applied Physics, University of Tokyo,
Bunkyo-ku, Tokyo 113, Japan}
\date{\today}
\maketitle
\begin{abstract}
 We study theoretically the phase diagram of 
perovskite manganites taking into account the double
degeneracy of the $e_g$ orbitals in a $\rm Mn^{3+}$ ion. 
A rich phase diagram is obtained in the mean 
field theory at zero temperature 
as functions of $x$ (hole concentration) and $J_S$
(antiferromagnetic interaction between $t_{2g}$ spins). 
 The global features of the phase diagram 
is understood in terms of the superexchange 
and double exchange interactions, which are 
strongly depends on types of the occupied $e_g$ orbitals.
The strong electron correlation induces the 
orbital polarization, which controls the 
dimension of the conduction band. 
A sequential change of the spin and orbital 
structures with doping holes is 
consistent with the recent experiments.
In particular, metallic A-type (layered) antiferromagnetic state is 
found for $x\sim0.5$ with the uniform $d_{x^2-y^2}$ orbital ordering.
In this phase, calculated results suggest the 
two-dimensional conduction and absence of the spin canting,
which are observed experimentally.
 Effects of the Jahn-Teller distortion are also studied.
\end{abstract}

\narrowtext
\section{Introduction}


 Doped manganites R$_{1-x}$A$_x$MnO$_3$ (R=La, Pr, Nd, Sm ;
A= Ca, Sr, Ba) have recently attracted considerable interests 
from the viewpoint of a close connection between the magnetism 
and the electric transport. \cite{chaha,hel,LaSr1,jin} 
 Theoretical studies of the double exchange interaction 
have been developed long time ago \cite{Zener,Anderson,deGenne} and 
explained the emergence of the ferromagnetism in doped 
manganites. \cite{Jonker}
 However recent systematic experimental studies have revealed more rich phase
diagrams in this system.
\begin{figure}[p]
\caption{}
\label{fig : Experiment}
\end{figure}
\noindent
 Observed phase diagrams of La$_{1-x}$Sr$_x$MnO$_3$ (wider band
width system) and of Pr$_{1-x}$Sr$_x$MnO$_3$ and  Nd$_{1-x}$Sr$_x$MnO$_3$ 
(narrower band width systems) are shown in Fig.\ref{fig : Experiment}.
\cite{LaSr1,PrSr,Kuwahara1}
 In the parent compound ($x=0.0$), the layered antiferromagnetism 
(spin A-type AF) accompanied with a 
distortion in the $\rm MnO_6$ octahedron 
is realized. \cite{Woll,Gen} 
 By moderately doping holes, the 
insulator with spin A-type AF changes into
an insulator with a ferromagnetism (spin F-type)
around $x=0.125$, and to
a spin F-type metal at $x \sim 0.175$. \cite{LaSr1,Okuda,Kawano1}
 In Pr$_{1-x}$Sr$_x$MnO$_3$ and Nd$_{1-x}$Sr$_x$MnO$_3$, 
a metallic phase with spin A-type AF 
is found recently to appear for $x>0.48$. \cite{PrSr,Kuwahara1,Kawano2} 
 With further increasing of $x$ ($x \sim 0.6$),
rod type antiferromagnetism (spin C-type AF) is found in 
La$_{1-x}$Ca$_x$MnO$_3$ \cite{Woll} and 
Nd$_{1-x}$Sr$_x$MnO$_3$. \cite{Kuwahara1}
 Finally, for $x=1.0$, the three dimensional AF (spin G-type AF) 
is realized. 
 In addition, in the narrower-band systems, the charge ordered state
accompanied by the spin ordering is 
recognized near the commensurate value of $x$ ($x$=0.5, 0.75 etc.), 
\cite{jirak,mori-} where the orbital is also supposed to be ordered. 
\par
In order to reveal the origin of 
the unique magnetotransport in this system, 
it is essential to understand the above rich 
phase diagram.
However, it cannot be explained by the conventional 
scenario based on the double exchange interaction.
 This discrepancy should be attributed to ingredients 
neglected in the conventional theory, namely the anisotropic transfer
intensity originated from the $e_g$ orbital degrees of freedom, 
electron-electron interactions,
and the electron-lattice interaction (Jahn-Teller (JT) effect).
 Especially, in the narrower-band systems, 
the kinetic energy of $e_g$ carriers is 
suppressed and the above interactions become more important.
 At $x=0.0$, the spin and orbital
structures have been studied theoretically 
by taking into account the above 
interactions since the pioneering 
works by Goodenough and Kanamori.\cite{Good,Kanamori} 
However, 
when we focus on the origin of the spin structure at $x=0$,
$i.e.$, the A-type AF, and roles of the JT distortion on it, 
situation is still controversial. 
In one side of the theoretical investigation, 
the spin ordering is attributed to the 
strong electron-electron interaction 
and the doubly degenerate $e_g$ 
orbitals. \cite{Kugel,Ishi1,Koshi,Shiina} 
The ferromagnetic superexchange interaction, 
which is necessary to explain the 
spin alignment in the $ab$-plane, 
is originated from the degenerate orbitals 
and the Hund coupling interaction between them. 
\cite{Good,Kanamori,Roth,Inag,Cyrot}
Even without JT effect, 
the A-ytpe spin alignment is derived 
by the superexchange interaction and the 
anisotropy of the transfer integral due to the orbital ordering. 
However, a type of the orbital ordering theoretically obtained 
disagrees with that expected from a type of the JT distortion, 
$i.e.$, $d_{3x^2-r^2}$ and $d_{3y^2-r^2}$. 
The another side of the theoretical investigation 
is based on the Hartree-Fock theory
\cite{Mizo} and the first-principle band calculation 
base on the local density approximation, \cite{Hamada} 
where the JT distortion is indispensable to reproduce the 
observed spin structure through introduction of the 
orbital polarization. 
Without the JT distortion, 
the system becomes a ferromagnetic metal. 
\par
 As for the doped case($x\ne0$), the situation is even more
controversial. 
 An issue of main interest here is the origin of the colossal 
magnetoresistance (CMR) observed near the ferromagnetic
transition temperature $T_c$.
 For this purpose identification of the dominant interactions
is of primary importance.
 Millis $et\ al.$ \cite{Millis} attributed an insulating behavior
of the resistivity $\rho(T)$ above $T_c$ to formation of the
small JT polaron. 
 They assumed the strong Hund's coupling, but the other
Coulomb interactions were neglected.
 The characteristic JT interaction is about $1eV$ which is comparable
or smaller than the band width.
 Then it is reasonable that the small polaron formation disappears 
below $T_c$ and $\rho(T)$
shows metallic behavior.
 However we argue that this picture does not explain the following
anomalous features observed in the low temperature ferromagnetic 
state ($T<\!\!<T_c$):
(i) $\rho(T)$ is fitted by
$\rho(T)=\rho_0 + A T^2$ with the coefficient $A$
being large, $i.e.$, 
$A \sim 500 \ \mu \Omega \ cm / K^2$. \cite{LaSr1} 
(ii) the optical conductivity $\sigma(\omega)$ is dominated by the
incoherent part with a small Drude weight. \cite{Okimoto}
(iii) photo-emission experiments show only a small discontinuity at the
Fermi edge. \cite{Sarma}
 We regard these features as signatures of the strong
correlation in the doped Mott insulator and the Coulomb
interactions remain strong even in the metallic state.
 Considering the strong Hund's coupling, which causes the perfect spin
polarization,
it is reasonable to assume the strong correlation between
$e_g$ electrons.
 In the ferromagnetic state, the orbital degrees of freedom play
a similar role to that of spins in the usual doped Mott insulator. 
It is the additional ingredient important in the manganites. 
\par
 In this paper we study a phase diagram of 
perovskite manganese oxides. 
The double degeneracy of the $e_g$ orbital, 
the anisotropy of the transfer integral based on it 
and the strong electron correlation are taken into account 
in the model. 
 The spin and orbital phase diagram 
at zero temperature is obtained by the mean field
approximation.
 The sequential change of the spin structure with doping of holes, 
$i.e.$, A $\rightarrow$ F $\rightarrow$ A $\rightarrow$ C $\rightarrow$ G, 
is well reproduced by the calculation. 
 We found that the strong Coulomb interactions experimentally suggested 
induce the perfect orbital polarization which plays
an essential role to determine the spin structure. 
 The orbital structure is also changed with doping of holes and 
controls the dimensionality of the conduction bands.
In particular, the metallic spin A-type AF phase 
is found around $x\sim0.5$ where the
orbitals are aligned as $d_{x^2-y^2}$.
In this phase, the interlayer electron transfer 
is forbidden both by the spin and orbital structures 
and the spin canting is absent. 
Both theoretical predictions are consistent with the recent 
experiments. \cite{PrSr,Kuwahara1}
Roles of the JT distortion in the undoped case 
is also studied. 
\par
 In Sec. II,  we introduce the model Hamiltonian and the formulation 
of the mean field calculation. 
Results of the numerical calculation are presented in Sec. III.
 Sect. IV is devoted to discussion and conclusion 
including comparison with the previous works.
A short version of this paper has been already published,
\cite{Maezono} but this paper contains additional and more detailed
results.
\section{Model and formulation}
 We start with the Hamiltonian
\begin{equation}
H=H_K+H_{Hund}+H_{on-site}+H_S \ ,
\label{eqn:eq1}
\end{equation}
where $H_K$ is the kinetic energy of $e_g$ electrons,
$H_{Hund}$ is the Hund coupling between $e_g$ and 
$t_{2g}$ spins, and $H_{on-site}$ represents the on-site
Coulomb interactions between $e_g$ electrons.
 $t_{2g}$ spins are treated as the localized spin with $S=3/2$.
 The AF coupling between nearest neighboring $t_{2g}$ spins 
is introduced in $H_S$ to reproduce the AF spin ordering observed at $x=1.0$.
 Using an operator $d_{i\sigma \gamma }^{\dagger}$ which creates
an $e_g$ electron with spin $\sigma$ ($={\uparrow}, {\downarrow}$)
in the  orbital $\gamma$ ($=a(d_{x^2-y^2}), b(d_{3z^2-r^2})$) at site $i$,
each term of Eq.(\ref{eqn:eq1}) is given by 
\begin{equation}
H_K=
\sum\limits_{\sigma \gamma \gamma' \langle ij \rangle} 
{t_{ij}^{\gamma \gamma '}d_{i\sigma \gamma }^{\dagger}d_{j\sigma \gamma '}}\ ,
\label{eqn : eq2}
\end{equation}
\begin{eqnarray}
H_{Hund}=-J_H\sum\limits_i {\vec S_{t_{2g} i}\cdot \vec 
S_{e_g i}^{}}\ , 
\label{eqn : eq4}
\end{eqnarray}
and 
\begin{equation}
H_S=J_S\sum\limits_{\left\langle {ij} \right\rangle } {\vec S_{t_{2g} i}
\cdot \vec S_{t_{2g} j}}\ .
\label{eqn : eq3}
\end{equation}
$ t_{ij}^{\gamma \gamma '}$ in $H_K$ is the 
electron transfer intensity between nearest neighboring sites 
and it depends on kind of occupied orbital and the direction of a bond as follows: \cite{Ishi2} 
\begin{equation}
  t_{i \ i+x}^{\gamma \gamma '}= t_0\left( {\matrix{{-{3 \over 4}}
     &{{{\sqrt 3} \over 4}}\cr
     {{{\sqrt 3} \over 4}}&{-{1 \over 4}}\cr
     }} \right) \ ,
     \\ 
\end{equation}
\begin{equation}
  t_{i \ i+y}^{\gamma \gamma '} = t_0\left( {\matrix{{-{3 \over 4}}
     &{-{{\sqrt 3} \over 4}}\cr
     {-{{\sqrt 3} \over 4}}&{-{1 \over 4}}\cr
     }} \right) \ ,
     \label{eqn : eq51}  
\end{equation}
and 
\begin{equation}
      t_{i \ i+z}^{\gamma \gamma '} = t_0\left( {\matrix{0&0\cr0&{-1}\cr
     }} \right) \ . 
     \end{equation}
$t_0$ is the electron transfer intensity between $d_{3z^2-r^2}$ orbitals 
along the $z$-direction. 
The spin operator for the $e_g$ electron is defined as 
$\vec S_{e_g i}={1 \over 2}\sum\limits_{\gamma \alpha \beta} 
 {d_{i\gamma \alpha }^{\dagger}\vec \sigma _{\alpha \beta }
d_{i\gamma \beta }^{}}$ with the Pauli matrices $\vec \sigma _{\alpha \beta }$. 
$\vec S_{t_{2g}i}$ denotes the localized $t_{2g}$ spin on $i$-site with $S=3/2$.
The last term in the Hamiltonian 
$H_{on-site}$ consists of the following three contributions:
\begin{equation}
H_{on-site}=H_U+H_{U'}+H_J \ ,
\label{eqn : eq5}
\end{equation}
 where $H_U$ and $H_{U'}$ are the intra- and the inter-orbital
Coulomb interactions, respectively, and $H_J$ denotes the inter-orbital
exchange interaction. 
 Each term is represented as
\begin{equation}
   H_U=U\sum\limits_{j\gamma } 
{n_{j\gamma \uparrow }}n_{j\gamma \downarrow } \ ,
\label{eqn : eq6} 
\end{equation}
\begin{equation}
H_{U'} =U'\sum\limits_{j\sigma \sigma '} {n_{ja\sigma }}n_{jb\sigma '} \ ,
\label{eqn : eq7} 
\end{equation}
and 
\begin{equation}
   H_J=J\sum\limits_{j\sigma \sigma '} {d_{
ja\sigma }^{\dagger}d_{jb\sigma '}^{\dagger}d_{ja\sigma '}^{}
d_{jb\sigma }^{}} \ ,
\label{eqn : eq8}
\end{equation}
with $n_{j\gamma \sigma }=d_{j\sigma \gamma }^{\dagger}
d_{j\sigma \gamma }^{}$ and $n_{j\gamma }=\sum\limits_\sigma
  {n_{j\gamma \sigma }}$. 
 Here, we assume that the three energy parameters are related as $U=U'+J$.
By using the spin operator for the $e_g$ electrons and the 
iso-spin operator describing the orbital degrees
of freedom, defined as 
\begin{equation}
\vec T_i={1 \over 2}\sum\limits_{\gamma \gamma' \sigma}  {d_{i\gamma \sigma }
^\dagger\vec \sigma _{\gamma \gamma '}d_{i\gamma '\sigma }^{}} \ ,
\label{eqn : eq9}
\end{equation}
$H_{on-site}$ can be rewritten by \cite{Ishi2}
\begin{equation}
H_{on-site}=-\sum\limits_i {\left( {\tilde \beta \vec T_i^2+\tilde
 \alpha \vec S_{e_{g} i}^2} \right)} \ .
\label{eqn:eq10}
\end{equation}
Coefficients of the spin and iso-spin operators, $i.e.$, 
$\tilde \alpha $ and $\tilde \beta $, are given by
\begin{equation}
\tilde \alpha = U-{J \over 2}>0\quad \ , 
\end{equation}
and 
\begin{equation}
\tilde \beta = U-{3J \over 2}>0 \ .
\label{eqn : eq11}
\end{equation}
The minus sign in Eq.(\ref{eqn:eq10}) means that the Coulomb 
interactions induce both spin and orbital (iso-spin) moments.
In the path integral representation,
the expression of the grand partition function
is represented as 
\begin{equation}
\Xi =\int { \prod\limits_{i}
D\vec S_{t_{2g} i}}D\bar d_{i\gamma \sigma }
  Dd_{i\gamma \sigma }^{}\exp \left\{ {-\int {d\tau \kern 
  1pt L\left( \tau  \right)}} \right\} \ ,
  \label{eqn : eq12} 
\end{equation}
with
\begin{equation}
  L\left( \tau  \right)=H\left( \tau  \right)+\sum\limits_
  {\sigma \gamma i} {\bar d_{\sigma \gamma i}\left( \tau  
  \right)\left( {\partial _\tau -\mu } \right)d_{\sigma 
\gamma i}}\left( \tau  \right)$$ \ ,
\label{eqn : eq13}
 \end{equation}
 where $\tau$ is the imaginary time
introduced in the path integral formalism, and
$ {\bar d_{i\gamma \sigma }^{}} $and $d_{i\gamma \sigma }^{}$ 
are the Grassmann variables corresponding to the
operators $ d_{i \gamma \sigma }^{\dagger}$ and $d_{i\gamma 
\sigma }^{} $, respectively. 
By utilizing Eq.(\ref{eqn:eq10}), 
Hamiltonian is rewritten by 
\begin{eqnarray}
  H\left( \tau  \right) & = & -\tilde \alpha \sum\limits_i {\left( 
   {\vec S_{e_g i}+{{J_H} \over {2\tilde \alpha }}\vec S_{t_{2g} i}}
   \right)^2} + {{J_H^2} \over {4\tilde \alpha }}\sum\limits_i
   {\vec S_{t_{2g} i}^2}\hfill\cr
  &   &  +J_S\sum\limits_{\left\langle {ij} \right\rangle }
   {\vec S_{t_{2g} i}\cdot \vec S_{t_{2g} j}}
   - \tilde \beta \sum\limits_i {\vec T_i^2}+H_K\ . 
   \label{eqn : eq14}
\end{eqnarray}
The bilinear terms with respect to the spin and iso-spin operators in the Hamiltonian 
is decoupled by introducing two kinds of auxiliary fields through  
the Stratonovich-Hubbard transformation. 
Then the partition function is rewritten as  
\begin{eqnarray}
  \Xi & = &  \int {  \prod\limits_{i}
    D\vec S_{t_{2g} i}}D\bar d_{i\gamma \sigma }Dd_{i\gamma
    \sigma }^{}D \vec{\varphi_{Si} } \vec{\varphi_{Ti} } e^{-\int{d\tau (L_{d}+L_{\vec \varphi})}}
  \label{eqn:eq15}    
\end{eqnarray}
with
\begin{eqnarray}
  L_{d} & = & \sum\limits_{\sigma \gamma i} {\bar d_{\sigma \gamma i}\left( 
          \tau \right)\left( {\partial _\tau -\mu } \right)d_{\sigma 
          \gamma i}}\left( \tau  \right)  \nonumber \\
 &+&\sum\limits_{\sigma \gamma \gamma ' \langle ij \rangle} 
         {t_{ij}^{\gamma \gamma '}\bar d_{\sigma \gamma i}(\tau)
         d_{\sigma \gamma 'j}(\tau)}  \nonumber \\
& - &   2\tilde \alpha \sum\limits_i
          {\vec S_{e_g i}(\tau)} \cdot {\vec \varphi _{Si}(\tau)} 
        -2 \tilde \beta \sum\limits_i 
  {\vec T_{i}(\tau)} \cdot {\vec \varphi _{Ti}(\tau)}    \ ,
\label{eqn:eq15.0}
\end{eqnarray}
and 
\begin{eqnarray}
   L_{\vec \varphi} & =  &  J_S \sum\limits_{\left\langle {ij} \right\rangle } 
          {\vec S_{t_{2g} i}(\tau)\cdot \vec S_{t_{2g} j}(\tau)}
          - J_H\sum\limits_i {\vec S_{t_{2g} i}(\tau)} 
            \vec\varphi_{S i}(\tau) \nonumber \\
    &   & + \tilde \alpha \sum\limits_i{\vec \varphi_{S i}^2(\tau)}
          + \tilde \beta \sum\limits_i {\vec \varphi_{T i}^2(\tau)}  \ . 
  \label{eqn:eq15.1}    
\end{eqnarray}
\par
Being based on the above formulae (Eq.(\ref{eqn:eq15})-(\ref{eqn:eq15.1})), 
we introduce the mean field approximation at this stage. 
 At first, we consider the part of the Hamiltonian which describes the 
$t_{2g}$ spin system, that is, the first and second terms in the right hand side 
in Eq. (\ref{eqn:eq15.1}). 
 The mean field solution in this system is given by
\begin{equation}
  {\left\langle {\vec S_{t_{2g i}}} \right\rangle } 
   =S {{\bar {\vec \phi _i}}\over {|\bar {\vec \phi _i}}|} \ ,
  \label{eqn : eq21}
\end{equation}
where $\bar {\vec \phi _i }$ is the solution of the following 
mean field equation
\begin{equation}
  \bar {\vec \phi _i}=-2J_S\sum\limits_j {}
{{S\bar {\vec \phi _j}}
  \over {\bar \phi _j}} + J_H\vec \varphi _{S i}^{} \ .
\label{eqn : eq22}
\end{equation}
By replacing the spin operator for the $t_{2g}$ spins 
by $\left\langle {\vec S_{t_{2g,i}}} \right\rangle$, 
$L_{\vec \varphi}$ in 
Eq. (\ref{eqn:eq15.1}) is given by 
\begin{eqnarray}
L_{\vec \varphi} & = & -J_H\sum\limits_i {\left\langle 
    {\vec S_{t_{2g} i}} \right\rangle }\vec \varphi 
     _{S i}(\tau)  \nonumber \\
  &   & +J_S\sum\limits_{ \langle ij \rangle} {\left\langle 
    {\vec S_{t_{2g} i}} \right\rangle \left\langle 
    {\vec S_{t_{2g} j}} \right\rangle }
  \nonumber \\
  &   & +\tilde \beta \sum\limits_i { \vec \varphi 
    _{T i}^2}+\tilde \alpha \sum\limits_i {\vec \varphi _{S i}^2} \ .
  \label{eqn : eq26}
\end{eqnarray}
Next, we focus on $L_{d}$ in Eq. (\ref{eqn:eq15.0}). 
By using the momentum representation: 
\begin{equation}
 d_{\sigma \gamma ' j}(\tau) = 
    {1 \over {\sqrt {\beta N}}}\sum\limits_k {\sum\limits_
    {n} {d_{\sigma \gamma '}(k,\omega_n)}}e^{i\vec k \cdot 
     \vec R_j-i\omega _n\tau } \ ,
 \label{eqn : eq27}
\end{equation}
and 
\begin{equation}
  \varphi _{x j}(\tau)  = {1 \over {\sqrt {\beta N}}}\sum\limits_k 
    {\sum\limits_{n} {\varphi _x(k,\omega_n)}}e^{i\vec k
    \cdot \vec R_j-i\omega _n\tau } \ ,
 \label{eqn : eq28}
\end{equation}
for $x=S$ and $T$, 
we have
\begin{eqnarray}
 \int {d\tau \kern 1pt L_d}\left( \tau  \right)
  \! &=&\!\!\!\!\sum \limits_{kk';nn'} {\sum\limits_{\gamma 
\gamma ';\alpha \beta } {\bar d_{\alpha \gamma}(k,\omega_n)}}
\nonumber \\
& \times &
  G_{kk';nn';\gamma \gamma ';\alpha \beta }^{-1}
  d_{\beta \gamma '}(k',\omega_{n'}) \ , 
 \label{eqn : eq2b}
\end{eqnarray}
where $\omega _n$ is the Matsubara-frequency for fermion 
and the $G_{kk';nn';\gamma \gamma ';\alpha \beta }$
is the Green function of the $e_g$ electron defined by 
\begin{eqnarray}
  G_{kk';nn';\gamma \gamma ';\alpha \beta }^{-1} & = & 
    \left( {-i\omega _n-\mu } \right)\delta _{nn'}\delta 
    _{\alpha \beta }\delta _{\gamma \gamma '}\delta _{kk'}
    \nonumber \\
  &+&  \varepsilon _k^{\gamma \gamma '}\delta _{kk'}\delta 
    _{nn'}\delta _{\alpha \beta }
    \nonumber \\
  &-&  {{\tilde \alpha } \over 
    {\sqrt {\beta N}}} \vec \sigma _{\alpha \beta } \cdot \vec\varphi 
    _S(k-k',\omega_n-\omega_{n'})\delta _{\gamma \gamma '}
    \nonumber \\
  &-& {{\tilde \beta } \over {\sqrt {\beta N}}} \vec \sigma
    _{\gamma \gamma '} \cdot \vec \varphi _T(k-k',\omega_n-\omega_{n'})
    \delta _{\alpha \beta } \ ,
  \label{eqn : eq30}
\end{eqnarray}
with   
\begin{equation}
  {1 \over N}\sum\limits_{\langle ij \rangle} {t_{ij}^{\gamma \gamma '}
  e^{-i\vec k\vec R_i+i\vec k'\vec R_j}}=\varepsilon 
  _k^{\gamma \gamma '}\delta _{kk'} \ .
\label{eqn : eq31}
\end{equation}
After integrating over the Grassman variable, 
the partition function is rewritten as 
\begin{eqnarray}
 \Xi & \!=\! & \int {D\left\{ \varphi  \right\}}\exp 
    \left( {Tr\ln G_{kk';nn';\gamma \gamma ';\alpha 
    \beta }^{-1}-\int {d\tau \kern 1pt L_{\vec \varphi}}} \right)
    \nonumber \\
  & \equiv & e^{-\beta \left( {F-\mu N} \right)} \ .
 \label{eqn : eq32}  
\end{eqnarray}
Then we adopt the mean field approximation 
by replacing the two kinds of auxiliary field, $i.e.$, 
$\vec \varphi_S$ and $\vec \varphi_T$ by their 
values at the saddle point 
$\bar {\vec \varphi_S}$ and $\bar{\vec \varphi_T}$. 
Finally, we obtain the expression for the free energy in the mean field 
approximation as 
\begin{eqnarray}
  F & = & \left. {L_{\vec \varphi}} \right|_{\left\{ {\varphi _x} \right\}
    =\left\{ {\bar \varphi _x} \right\}}
    \nonumber \\
  & & -{1 \over \beta }\sum\limits_\nu  {\ln \left[ {1+\exp \left
    \{ {-\beta \left( {E^{\left( \nu  \right)}-\mu } \right)} \right\}}
    \right]}_{\left\{ {\varphi _x} \right\}=\left\{ {\bar \varphi _x} 
    \right\}}
    \nonumber \\
  & & +\mu N \ ,
\label{eqn : eq35}  
\end{eqnarray}
where $E^{(\nu)}$ is the $\nu$-th eigenvalue of 
$M_{kk';\gamma \gamma ';\alpha \beta }$ defined by 
\begin{eqnarray}
  M_{kk';\gamma \gamma ';\alpha \beta } & = & \varepsilon 
  _k^{\gamma \gamma '}\delta _{kk'}\delta _{\alpha 
  \beta }-{{\tilde \alpha } \over {\sqrt N}}\vec 
  \sigma _{\alpha \beta } \cdot \vec\varphi _S(k-k')
  \delta _{\gamma \gamma '} \nonumber \\
 &   & -{{\tilde \beta } \over {\sqrt N}}\vec 
  \sigma _{\gamma \gamma'} \cdot \vec\varphi _T(k-k')
  \delta _{\alpha \beta } \ . 
\label{eqn : eq37}  
\end{eqnarray}
Chemical potential $\mu$ is determined by the following condition 
\begin{equation}
  \left( {1-x} \right)={1 \over N}\sum\limits_\nu  
  {f\left( {E^{\left( \nu  \right)}-\mu } \right)} \ ,
\label{eq36}  
\end{equation}
in terms of the doping concentration $x$. 
$f\left( x \right)$ 
is the Fermi distribution function.
\par
By using the above expression of the free energy, 
we numerically calculate the spin and orbital phase diagram 
at zero temperature. 
 We consider four kinds of spin alignment in the cubic cell: F-, 
A-, C- and G-type. The possibility of the spin canting is also 
discussed later.
 As for the orbital degrees, their ordering is represented by the alignment of
the iso-spin. 
We specify the orbital state by the angle in the orbital space as follows:
\begin{eqnarray}
  \left| \theta  \right\rangle  
  & = &  \cos {\theta \over 2} \left|d_{x^2-y^2} \right\rangle 
       + \sin {\theta \over 2}  \left| d_{3z^2-r^2} \right\rangle \ ,
\label{eqn : eq40}
\end{eqnarray}
which describes the direction of the iso-spin moment 
\begin{equation}
  \vec T = (-\sin \theta,0,\cos \theta) \ .
\label{eqn : eq40-1}
\end{equation}
 We also consider four types of orbital ordering, $i.e.$, F-,
A-, C-, G-type, in the cubic cell. 
 The angle in the orbital space $\theta$ is varied for each 
sublattice, and these are denoted as $\theta_{I}$ and $\theta_{II}$ 
in the $I$ and $II$ sublattices, respectively. 
 Henceforth, we often use the notation such as,
orbital $G:\,{{\left( {3x^2-r^2} \right)} 
\mathord{\left/ {\vphantom {{\left( {3x^2-r^2} \right)}
{\left( {3y^2-r^2} \right)}}} \right. \kern-
\nulldelimiterspace} {\left( {3y^2-r^2} \right)}}$ = $\left(G:\,
{{\pi  \mathord{\left/ {\vphantom {\pi  3}} \right. 
\kern-\nulldelimiterspace} 3},{{\,-\pi } \mathord{\left/ 
{\vphantom {{\,-\pi } 3}} \right. \kern-\nulldelimiterspace} 
3}} \right)$, through the relations,
\begin{equation}
  \left|d_{3x^2-r^2} \right \rangle  =  \cos {{\left( {{\pi  
    \mathord{\left/ 
    {\vphantom {\pi  3}} \right. \kern-\nulldelimiterspace} 3}}
    \right)} \over 2} \left|d_{x^2-y^2} \right\rangle
    +\sin {{\left( {{\pi \mathord{\left/ {\vphantom {\pi 3}} \right. 
    \kern- \nulldelimiterspace} 3}} \right)} \over 2} 
    \left| d_{3z^2-r^2} \right\rangle\ ,
\end{equation}
and 
\begin{equation}
  \left|d_{3y^2-r^2}\right\rangle  =  \cos {{\left( {{{-\pi } 
    \mathord{\left/ 
    {\vphantom {{-\pi } 3}} \right. \kern-\nulldelimiterspace}
     3}} \right)} \over 2}  \left|d_{x^2-y^2} \right\rangle
    +\sin {{\left( {{{-\pi } \mathord{\left/ {\vphantom 
    {{-\pi } 3}} \right. \kern-\nulldelimiterspace} 3}} \right)} 
    \over 2}  \left| d_{3z^2-r^2} \right\rangle \ .
\end{equation}
 Therefore, we consider the 4 (spin) $\times$ 4 (orbital) types
of ordering with ($\theta_I, \theta_{I\!I}$),
and numerically compare the free energy between them. 
%
\section{Numerical Results}
\subsection{Parameters in the model}
 The values of the energy parameters $\tilde \alpha ,\tilde \beta ,J_H,J_S,t_0$, 
used in the numerical calculation are chosen as follows.
 In LaMnO$_3$, $t_0$ is estimated by the photoemission 
experiments to be $t_0 \sim 0.72$eV, \cite{Saitoh} 
which we choose the unit of the energy below ($t_0=1$).
 By employing $U=6.3$eV and $J = J_H=1.0$eV as those 
relevant to the actual manganese oxides, \cite{Saitoh} 
parameters for the electron-electron interactions in the present model 
are $\tilde \alpha=8.1,\  \tilde \beta=6.67$ \ 
($\tilde \alpha / \tilde \beta =1.21$).
 The numerical calculation are also performed 
by using different sets of the energy parameters 
in order to compare the previous works. \cite{Kugel,Ishi1,Koshi,Shiina}
There, the effective Hamiltonians are derived by excluding the doubly 
occupied state in the $e_g$ orbials. 
The superexchange interaction between nearest neighboring spin and orbital 
in these models are characterized by the energy in the intermediate states in the perturbation process. 
There are three kinds of intermediate states,
\cite{Roth,Inag,Cyrot,Kugel} $i.e.$, the two electrons occupy 1) 
the different orbitals with the parallel spin (the energy is $U'-J$), 2) 
the different ones with the antiparallel spins ($U'+J$), 
and 3) the same orbital ($U$).
The connection between these energies and the present energy parameters 
are roughly estimated as $U'-J \sim \tilde \beta$,\ $U \sim \tilde \alpha$, and $U'+J \sim 
\tilde \alpha +\tilde \beta$ from Eq.(\ref{eqn:eq10}).
Koshibae $et\ al$. \cite{Koshi} discussed the orbital 
ordering at $x\!=\!0$ by using the exact diagonalization method 
in the limit of $U'-J <\!< U, \ U'+J $ corresponding to 
$\tilde \alpha / \tilde \beta >\!>1 $. 
Shiina $et\  al.$ \cite{Shiina} also studies the spin and orbital structure 
in the wide range of the parameters.  
In order to compare the above results, 
we show the two cases, that is, 
(case (A)) with $\tilde \alpha=70$ and  
$\tilde \beta=2.5$ ($\tilde \alpha / \tilde \beta >\!> 1$), and 
(case (B)) with $\tilde \alpha =8.1$ and $\tilde \beta =6.67$.
\par
\subsection{Undoped ($x=0$) case}
Let us consider the undoped case. 
\begin{figure}[p]
\caption{}
\label{fig : Free}
\end{figure}
\noindent
In Fig.\ref{fig : Free}, we show the 
calculated free energy $\!F\left( {J_S},x=0 \right)$ 
at $x=0$ as a function of $J_S$ 
for each spin alignment in the case of $\tilde \alpha / \tilde \beta >\!>1$ 
(case (A)). 
In each spin alignment, the orbital structure are optimized. 
 In the case of $J_S=0$, the F-type spin alignment are favored 
due to the ferromagnetic superexchange interaction under 
the doubly-degenerate orbital. \cite{Roth,Inag,Cyrot} 
 With increasing $J_S$, the stable spin structure is changed 
from F- to A- , C- and finally G-type. 
This sequential change of the spin structure is consistent 
with the theoretical studies based on the effective Hamiltonian 
\cite{Ishi1,Koshi} in the limit $\tilde \alpha / \tilde\beta >\!> 1$.
\begin{figure}[p]
\caption{}
\label{fig : SOdep}
\end{figure}
\noindent
 Spin and orbital structures are depend on 
$J_S$ and these are depicted in Fig.\ref{fig : SOdep}. 
 Rearrangement of the orbital structure with changing $J_S$, 
which is previously pointed out, \cite{Koshi} 
are also found. 
\par
 It is worth noting that the orbital structures 
also depend on the ratio $\tilde \alpha / \tilde \beta$.
\begin{figure}[p]
\caption{}
\label{fig : thetadep}
\end{figure}
\noindent
In Fig.\ref{fig : thetadep}, 
the angle in iso-spin space $\left( 
{\theta _I,\theta _{I\!I}} \right)$ obtained in the spin A phase 
is presented. 
For $\tilde \alpha / \tilde \beta >\!>1$ the configuration
$\left( {\theta _I,\theta _{I\!I}} \right)=(90,-90)$ is obtained as 
the stable orbital structure. 
In this orbital ordering, the energy gain due to the super exchange 
process in the ferromagnetic bonds, $i.e.$, 
${{t^2} \mathord{\left/ {\vphantom {{t^2} {\left( {U'-J} 
\right)}}} \right. \kern-\nulldelimiterspace} 
{\left( {U'-J} \right)}}$ takes its maximum.
Here, $t$ represents the effective electron transfer intensity in this 
superexchange process including the effects of orbital. 
On the other hand for $\tilde \alpha / \tilde \beta<\!<1$,
the superexchange processes in the AF bonds characterized by 
${{t^2} \mathord{\left/ {\vphantom {{t^2} U}} \right. \kern-
\nulldelimiterspace} U}$ and ${{t^2} \mathord{\left/ {\vphantom 
{{t^2} {\left( {U'+J} \right)}}} \right. \kern-\nulldelimiterspace} 
{\left( {U'+J} \right)}}$, become important.
 Actually $\left( {\theta _I,\theta
_{I\!I}} \right)=(180,-180)$ is the most preferable
configuration in this parameter region. 
 Orbital $G\!\!:\!\!{\left( {3x^2-r^2} \right)}/
{\left( {3y^2-r^2} \right)}$ ($\left( {\theta _I,\theta
_{I\!I}} \right)\!\!=\!\!(60,-60)$), 
which is supposed experimentally, 
can not be the most stable for any $\tilde \alpha /\tilde \beta$.
Rather than this structure, $G\!\!:\!\!{{\left( {y^2-z^2} \right)} 
\mathord{\left/ {\vphantom {{\left( {y^2-z^2} \right)} 
{\left( {x^2-z^2} \right)}}} \right. \kern-\nulldelimiterspace} 
{\left( {x^2-z^2} \right)}}$ ($\left( {\theta _I,
\theta _{I\!I}} \right)\!\!=\!\!(120,-120)$) becomes stable around 
$\tilde \alpha / \tilde \beta \sim 1.1$.
\begin{figure}[p]
\caption{}
\label{fig : OrbforA}
\end{figure}
\noindent
 These numerical results are understood by comparing the energy 
gains due to the superexchange processes between the two orbital 
configurations shown in Fig. \ref{fig : OrbforA}. 
In the processes with the energy of $t^2/(U'+J)$ and 
$t^2/ {\left( {U'-J}\right)}$,
the transfer integral $t$ between the occupied 
and the unoccupied orbital is concerned and it 
takes the same value for both configurations. 
 On the other hand, in the process with $t^2/{\left( U \right)}$,
relevant transfer is the one between the occupied orbitals 
along $c$-axis, which is always larger for orbital: 
$G\!\!:\!\!{\left( {y^2-z^2} \right)}/{\left( {x^2-z^2} \right)}$ 
than that for orbital:$G\!\!:\!\!{\left( {3x^2-r^2} \right)}/
{\left( {3y^2-r^2} \right)}$.
 Then there is no chance for orbital $G\!\!:\!\!{{\left( {3x^2-r^2} 
\right)} \mathord{\left/ {\vphantom {{\left( {3x^2-r^2} \right)} 
{\left( {3y^2-r^2} \right)}}} \right. \kern-\nulldelimiterspace} 
{\left( {3y^2-r^2} \right)}}$ to be the most stable structure 
for any $\tilde \alpha / \tilde \beta $, when only the super exchange 
mechanism is considered. 
 Hence we conclude that the JT coupling plays an indispensable 
role for the $G\!\!:\!\!{{\left( {3x^2-r^2} 
\right)} \mathord{\left/ {\vphantom {{\left( {3x^2-r^2} \right)} 
{\left( {3y^2-r^2} \right)}}} \right. \kern-\nulldelimiterspace} 
{\left( {3y^2-r^2} \right)}}$ orbital ordering at $x=0$, which we will discuss
in the next section.
\par
In the vicinity of $\tilde\alpha/\tilde\beta\!=\!1$, 
A-type spin ordering appears even at $J_S=0$.
 This is consistent with the result obtained in the effective 
Hamiltonian at $x=0$. \cite{Shiina} 
 In this region, the ferromagnetic interaction 
characterized by the energy 
${t^2}/{\left( {U'-J} \right)}$ is suppressed, 
on the other hand, the AF ones with ${t^2}/{U}$ and 
${t^2}/{\left( {U'+J} \right)}$ are enhanced. 
When $\tilde\alpha/\tilde \beta$ approaches to unity 
with fixing $J_H$ as several eV, the spin F structure at 
$J_S=0$ remains. 
It is concluded that the strong $J_H$ plays an 
important role for the emergence of spin F at $J_S=0$. 
%
%
\subsection{Effects of the lattice distortion}
With considering the experimental fact that 
the static JT distortion rapidly disappears around 
$x\sim0.1$, \cite{Kawano1,Endoh}
we examine the effect of the JT distortion on the spin and orbital 
phase diagram at $x=0$.
 The JT distortion directly affects
the orbital configuration. 
As a result, 
the phase boundary between spin F and spin AF phases is modified. 
 To examine how the phase boundary is changed due to the JT distortion, 
we chose the parameters as $\tilde\alpha/\tilde\beta >\!>1$, $i.e.$, case (A), 
where the phase boundary exists at $x=0$ as shown in Fig. 9. 
The JT coupling is expressed as the coupling 
between the iso-spin 
operator $\vec T_j$ and the local lattice distortion $\vec Q_j$ as follows: \cite{Kanamori}
\begin{eqnarray}
  H_{JT} & = & g \sum\limits_j {\vec Q_j \cdot \vec T_j}
            + \frac{1}{2}\sum\limits_j {\omega_0^2\vec Q_j^2}
            + \sum\limits_j {V(\vec Q_j)} \ ,
  \\
\label{eqn : eq54}
\end{eqnarray}
 where $V(\vec Q_j)$ is the anharmonic potential for $\vec Q_j$.
 Instead of minimizing the total energy, we assume $\vec Q_j$ 
as the experimentally observed one.
 In a MnO$_6$ octahedron, $\vec Q_j^{}$ is expressed as  
\begin{equation}
  \vec Q_j=r_j \left( {\sin \Theta _j \hat Q_{x^2-y^2}
   +\cos \Theta _j \hat Q_{3z^2-r^2}} \right) \ ,
\label{eqn : eq56}
\end{equation}
 where $\hat Q_{x^2-y^2(3z^2-r^2)}$ is the base of the normal coordinate of 
the cubic-symmetric system defined as 
$\hat Q_{x^2-y^2}\!=\!{1 \over {\sqrt 2}}\left( {\Delta _x-\Delta _y} 
\right)$ and $\hat Q_{3z^2-r^2}\!=\!{1 \over {\sqrt 6}}\left( {2\Delta _z-
\Delta _x-\Delta _y} \right)$, and $\Delta _\alpha $ denotes the elongation 
toward the $\alpha$-direction ($\alpha = x, y, z$). 
 In this notation, the first term in $H_{JT}$, 
which is termed $H_{el-ph}$, is expressed as,
\begin{equation}
  H_{el-ph}=+\left| g \right|\sum\limits_j {r_j\vec v_j \cdot \vec T_j} \ ,
\label{eqn : eq57}
\end{equation}
with 
\begin{equation}
  \vec v_j=\left( {\matrix{{\sin \Theta _j}\cr0\cr
  {\cos \Theta _j}\cr}} \right) \ .
\label{eqn : eq58}
\end{equation}
We choose the sign of the coupling constant $+\left|g\right|$ so that 
the $d_{3z^2-r^2}$ orbital may be stabilized for 
$\hat Q_{3z^2-r^2}$, consistent with the negative charge of oxygen ion. 
 By the X-ray diffraction experiment, it is confirmed that 
the MnO$_6$ octahedron is elongated along
the $x$- or $y$-direction and these octahedron are alternatively aligned in the $ab$-aplane. \cite{Gen} 
In the present formula, it corresponds to $r_j=r$, $\Theta _j=-60^\circ$ for $j \in I$-sublattice and
$\Theta _j=-60^\circ$ for $j \in I\!I$-sublattice, $i.e.$, ($\Theta_I,\Theta_{I\!I}$)=
($60^\circ,-60^\circ$).
 By adding $H_{el-ph}$ to the Hamiltonian 
introduced in the previous section (Eq.(\ref{eqn:eq1})), 
an additional molecular field for the iso-spin is introduced. 
As a result, 
the matrix element $M_{kk';\gamma \gamma ';\alpha \beta }$
in Eq.(\ref{eqn : eq37}) is modified as
\begin{eqnarray}
  \lefteqn{M_{kk';\gamma \gamma ';\alpha \beta }\ }
     \nonumber \\
  & = &\varepsilon _k^{\gamma \gamma '}\delta _{kk'}\delta 
     _{\alpha \beta }-\tilde \alpha \vec \sigma _{\alpha \beta }
     \varphi _S^{}\delta _{ k- k'+ q_S}\delta _{\gamma \gamma '}
     \nonumber \\
  &   & -\tilde\beta \left \{
       \left( {\varphi _T^A
          -{{\left| g \right|r} \over {2\tilde \beta }}v^A} \right)\delta 
            _{k-k'}  \right. \nonumber \\
    \hskip 2cm  &+& \left. \left( {\varphi _T^C
          -{{\left| g \right|r} \over {2\tilde \beta }}v^C} \right)\delta 
            _{k-k'-q_\theta}
     \right \} (\sigma_z) _{\gamma \gamma '} \delta _{\alpha \beta}
     \nonumber \\
  &   & -\tilde\beta \left \{
       \left( {-\varphi _T^B
          -{{\left| g \right|r} \over {2\tilde \beta }}v^{BD}}\right)\delta 
            _{k-k'} \right. \nonumber \\
   \hskip 2cm   &+& \left. \left( {\varphi _T^B
          -{{\left| g \right|r} \over {2\tilde \beta }}v^{BA}}\right)\delta 
            _{k-k'-q_\theta}
     \right \} (\sigma_x)_{\gamma \gamma '} \delta _{\alpha \beta}
\label{eqn : eq59}
\end{eqnarray}
where
\begin{eqnarray*}
 v^{BD} &=& -\cos \left( {-120^\circ} \right)\sin \theta _I \ , \\ 
 v^{BA} &=& \sin \left( {-120^\circ} \right)\cos \theta _I \ , \\
 v^A &=& \cos \left( {-120^\circ} \right)\cos \theta _I \ , \\
  v^C &=& \sin \left( {-120^\circ} \right)\sin \theta _I \ , \\
  \varphi^A_T &=& \varphi_T \sin^2 \left( {\frac{\theta_I -\theta_{I\!I}}{2}
   } \right) \ ,\\
  \varphi^B_T &=& \varphi_T \cos \left( {\frac{\theta_I -\theta_{I\!I}}{2}
   } \right) \sin \left( {\frac{\theta_I -\theta_{I\!I}}{2}
   } \right) \ , \\
  \varphi^C_T &=& \varphi_T \cos^2 \left( {\frac{\theta_I -\theta_{I\!I}}{2}
   } \right) \ ,
\end{eqnarray*}
and $q_S$ ($q_\theta$) denotes the wave vector for the spin (orbital)
ordering.
 As well as the energy splitting between the two $e_g$ orbital 
due to the JT distortion (Eq.(\ref{eqn : eq57})), 
the distortion modifies the transfer integral 
through the modification of the bond length $l$ between Mn and O ions.
 According to the pseudo potential theory, \cite{Harrison} 
the overlap integral between Mn 3$d$ and 
O 2$p$ orbitals is proportional to $l^{{7 \mathord{\left/ {\vphantom 
{7 2}} \right. \kern-\nulldelimiterspace} 2}}$. 
 Therefore, the variations of the transfer 
integral between Mn 3$d$ orbitals is evaluated by using the parameter 
$r (=r_j)$ as
\begin{equation}
  t^{\gamma \gamma '}_{i \ i+{\hat x}({\hat y})}\left( r \right) = {{t^{\gamma 
     \gamma '}_{i i+{\hat x}({\hat y})}\left( {r=0} \right)} \over {\sqrt {\left( {1+2r} 
     \right)^7\left( {1-r} \right)^7}}} \ ,
\label{eqn : eq61a}
\end{equation}
and 
\begin{equation}
  t^{\gamma \gamma '}_{i i+{\hat z}}\left( r \right) =  {{t^{\gamma \gamma '}_{i i+{\hat z}}
     \left( {r=0} \right)} \over {\left( {1-r} \right)^7}} \ ,
     \label{eqn : eq61b}  
\end{equation}
where we used the expression for the bond lengths in the elongated 
and shorten bonds as 
$ l_{long}=l_0\left( {1+2r} \right)$ and 
$ l_{short}=l_0 \left( {1-r} \right)$, respectively. 
We also consider the change in the magnitude of $J_S$ 
due to the JT distortion. 
 With the relation $J_S^{}\left( r \right)\propto t_{}^2
\left( r \right)$, following relations are derived
\begin{equation}
  J_S^{x,y}\left( r \right)={{J_S^{}\left( {r=0} \right)} 
    \over {\left( {1+2r} \right)^7\left( {1-r} \right)^7}} \ , 
\end{equation}
and
\begin{equation} 
J_S^z\left( r \right)={{J_S^{}\left( {r=0} 
    \right)} \over {\left( {1-r} \right)^{14}}} \ , 
   \label{eqn : eq63}    
  \end{equation}
where $J_S^{\alpha} (r)$ is the superexchange interaction along 
$\alpha$ direction with the distortion $r$. 
 According to the X-ray diffraction experiment 
,\cite{Gen} bond lengths are reported as 
$  l_{long}^{} = 2.14 \AA$ and $l_{short}^{}=$ 1.98 or 1.96 $\AA$
corresponding to $r\!=\!0.028$ . 
 In order to distinguish the two kinds of modification due to the JT distortion, that is, 
the energy level splitting and the modification of the transfer intensity, 
we examine these effects separately by two procedures as 
changing the value of $g$ with fixing $r$ and vice versa. 
Even for $g=0$, the modification of the transfer intensity lifts the degeneracy.
\par
\begin{figure}[p]
\caption{}
\label{fig : JT-1}
\end{figure}
In Fig.\ref{fig : JT-1}, the stable orbital structure is shown as 
a function of the diagonal coupling $gr$ with fixing $r$.  
 For $g\!=\!0$,
($\theta _I$, $\theta _{I\!I}$) is 
determined so as to lower the center of mass in the valence band. 
 For sufficiently large $g$ compared with $t_0$,
JT distortion forces the orbital configuration to be 
($\theta _I$,$\theta _{I\!I}$)=($60,-60$).
 Types of the stable orbital are almost saturated as 
${{\left( {3x^2-r^2} \right)} \mathord{\left/ {\vphantom {{
\left( {3x^2-r^2} \right)} {\left( {3y^2-r^2} \right)}}} \right. \kern-
\nulldelimiterspace} {\left( {3y^2-r^2} \right)}}$ for
${{gr}/{t_0}}\!>\!0.5$.
  When the Coulomb interactions are not introduced,
this value $gr/t_0 \sim 0.5$ is not enough to make the wave 
functions to be ${{\left( {3x^2-r^2} \right)} \mathord{\left/ 
{\vphantom {{\left( {3x^2-r^2} \right)} {\left( {3y^2-r^2} \right)}}} 
\right. \kern-\nulldelimiterspace} {\left( {3y^2-r^2} \right)}}$.
 Since the orbital is already polarized by the strong Coulomb interaction, 
the role of JT coupling is to rotate
the direction of its polarization. 
It is much easier than
to induce the polarization. 
We concluded that the wave function is almost ($3x^2-r^2$)/($3y^2-r^2$) at $x=0$, 
which in principle can be tested experimentally. \cite{Murakami,Ishi3}
\par
\begin{figure}[p]
\caption{}
\label{fig : JT-2}
\end{figure}
 In Fig.\ref{fig : JT-2}, the variation of the phase boundary 
between spin F- and spin A-phases are presented as functions 
of $r$ and $gr$. 
The value of the superexchange interaction at the boundary is 
termed $J_S(FA)$, hereafter.
In the case of $g=0$ (Fig. \ref{fig : JT-2} (a)), 
spin F is stabilized with increasing $r$. 
 This is reasonable because the modification of the transfer 
intensity, described by Eqs. (\ref{eqn : eq61a}) and (\ref{eqn : eq61b}),
enhances the ferromagnetic superexchange 
interaction along the $c$-axis. 
On the other hand, introducing of 
the energy splitting represented by $gr$ with fixing $r$ stabilizes 
the spin A structure (Fig. \ref{fig : JT-2} (b)).
In order to understand this results,  
we estimate the energy gain due to the energy splitting as follows. 
The energy difference between the spin F and spin A phases are represented 
as 
\begin{equation}
2 S^2  J_S\left( {FA} ;g \right)=E_{A} (g)-E_{F}(g)
\label{eqn : eq65}  
\end{equation}
where $E_{A(F)}(g)$ is the energy for the spin-F(A) with the 
JT coupling $g$, and $J_S\left( {FA} ;g \right)$
is the superexchange interaction at the phase boundary. 
The prefactor 2 in the left hand side is came from 
difference of the number of the antiferromagnetic bonds between 
two phases. 
By using this expression, the change of 
the phase boundary between the $g=0$ and $g=\infty$ is 
estimated by
\begin{eqnarray}
  \lefteqn{2 S^2\left( {J_S \left( {FA};g=0 \right)-J_S\left( {FA};g=\infty 
     \right)} \right)} \nonumber \\
& = & \left( {E_{A}(g=0)-E_{A}(g=\infty) } \right) \nonumber \\
&-&  \left( {E_{F}(g=0)-E_{F}(g=\infty) } \right) \ .
\label{eqn : eq66}
\end{eqnarray}
When the right-hand side of Eq.(\ref{eqn : eq66}) is positive, 
the phase boundary $J_S^{}\left( {FA;g} \right)$ is increased 
with decreasing $g$.
Because the ferromagnetic superexchange interaction is only relevant at $x=0$ 
with the condition $\tilde \alpha / \tilde \beta >\!>1$, 
$ E_{F(A)}(g)$ is proportional to the sum of square of the transfer intensity 
between the nearest neighboring occupied and unoccupied orbitals 
$(t_{o-u}(g))$, that is, 
\begin{equation}
 E_{F (A)}(g) \propto  {-\sum\limits_{Ferro-bonds} {t_{o-u}
^2\left( {g} \right)}}  \ , 
\label{eqn : eq67}  
\end{equation}
where $\sum\limits_{Ferro-bonds}$ implies the summation 
over the ferromagnetic bonds. 
\begin{figure}[p]
\caption{}
\label{fig : JT-4}
\end{figure}
\noindent
These quantities are tabulated in Fig.\ref{fig : JT-4}. 
As a result, we obtain 
\begin{eqnarray}
\lefteqn{2S^2\left( {J_S\left( {FA;g=0} \right)-J_S \left( {FA;g=\infty} 
\right)} \right)}
    \nonumber \\
    & \propto & -\left( {I_{xy}^{A,g=0}-I_{xy}^{A,g=\infty}} \right)
    \nonumber \\
    & + & \left\{ 
    {\left( {I_{xy}^{F,g=0}-I_{xy}^{F,g=\infty}}
    \right) 
     + \left( {I_{z}^{F,g=0}-I_{z}^{F,g=\infty}} \right)} 
   \right \} \nonumber \\
   & = & 0.875>0 \ , 
\label{eqn : eq68}  
\end{eqnarray}
with
\begin{eqnarray}
 I_{xy} 
   &=& \frac{1}{2} \left\{ \sum (t_{o-u}^x)^2+\sum (t_{o-u}^y)^2 \right\}
    \nonumber  \ , \\
 I_{z} &=& \frac{1}{2} \left\{ \sum (t_{o-u}^z)^2 \right\}
    \nonumber \ .
\label{eqn : eq68-2}  
\end{eqnarray}
We conclude that $J_S^{}\left( {FA;g} \right)$ increases
with decreasing $g$ as shown in Fig.\ref{fig : JT-2} (b). 
This results implies that 
with relaxing the JT distortion, 
a frozen of the orbital configuration is melted from 
($\theta _I$, $\theta _{I\!I}$) $=$ ($60,-60$), 
the spin F phase is stabilized in comparison with spin A phase. 
It is consistent with the experimental results where the spin-A phase is 
replaced by the spin-F insulator accompanied with reduction of the 
JT distortion as increasing $x$, although the present 
calculation is limited in the undoped case. 
%
%
\subsection{Doped ($x\ne 0$) case}
In this subsection, we show the results in the finite hole doped case.
\begin{figure}[p]
\caption{}
\label{fig : PD28}
\end{figure}
\begin{figure}[p]
\caption{}
\label{fig : PD1.21}
\end{figure}
\noindent
 In Fig.\ref{fig : PD28} and \ref{fig : PD1.21}, 
we present the phase diagrams as a function of 
hole concentration $x$ in the cases of (A) ($\tilde\alpha=70, \tilde\beta=2.5$)
and (B) ($\tilde \alpha =8.1, \tilde \beta =6.67$), respectively.
 Nonmonotonic behavior of the phase boundary is 
attributed to changes of the orbital structures.
In both cases, the global features of phase diagram 
are almost the same.
 As discussed before, $\tilde \alpha / \tilde \beta >\!>1$ 
corresponds to $U'-J <\!< U, \ U'+J $ and the large Hund coupling. 
 In this limit, the superexchange process for the AF interaction 
is neglected and the analyses of the calculated results become easier. 
Therefore, at first, we focus on the results in the case (A). 
\par
It is found that 
in nearly the whole doping region except for $x \cong 0$,
$\left( {3z^2-r^2} \right)$
and $\left( {x^2-y^2} \right)$ orbital structures are 
stabilized in the spin A- and C-type phases, respectively, 
because these orbitals are favorable to maximize a gain of the kinetic energy 
in each spin structure. 
 Deviation from $\left( {x^2-y^2} \right)$ structure in the A-type AF phase 
is found in $0.45\!<\!x\!<\!0.75$ and it is attributed to the 
hybridization between the occupied and unoccupied bands as 
discussed later.
 For spin G-type AF, the energy does not depend on 
the orbital so much, because the electron motion is blocked 
in all directions. 
In the spin F structure, on the other hand, the transfer is allowed 
in any direction.
 Orbital structure in spin F changes continuously as $x$ 
increases from orbital $G\!\!:\!\!{{\left( {x^2-y^2} \right)}/
{\left( {3z^2-r^2} \right)}}$ 
near $x=0$ to orbital F: $\left( {x^2-y^2} \right)$ for $x \sim 0.3$, and to orbital 
$A\!\!:\!\!{{\left( {\left[ {3z^2-r^2} \right]+\left[ {x^2-y^2} \right]} 
\right)}/
{\left( {\left[ {3z^2-r^2} \right]-\left[ {x^2-y^2} \right]} \right)}}$ 
for $0.3\!<\!x\!<\!0.8$ and finally orbital F: $\left( {3z^2-r^2} \right)$. 
 The orbital structure is sensitively changed by changing $x$ in comparison 
with that in the other spin structures.
\par
 In order to understand the variation of the spin and orbital structures 
in the finite hole concentration region, 
let us consider the density of states (DOS) for each orbital configuration.
\begin{figure}[p]
\caption{}
\label{fig : band-str}
\end{figure}
\noindent
 In Fig.\ref{fig : band-str}, we present the 
schematic picture of DOS for several 
values of $\tilde \alpha$ and $\tilde \beta$. 
In the case of large $\tilde\alpha$ and $\tilde\beta$, 
the density of states is split into four bands. 
Each band are characterized by the direction of the 
spin and iso-spin, that is, the spin 
is parallel or anti-parallel and the iso-spin is parallel or anti-parallel 
to their mean fields, respectively.
It accommodates an electron per site, and the lowest band 
corresponds to the state where both spin and iso-spin are parallel. 
 Energy difference between the two bands 
where the (iso)spin are parallel and anti-parallel, respectively, 
is given by $\tilde \alpha \varphi _S{{=\tilde \alpha \left( {1-x} 
\right)} \mathord{\left/ {\vphantom {{=\tilde \alpha \left( {1-x} \right)}
 2}} \right. \kern-\nulldelimiterspace} 2}$ 
$(\tilde \beta \varphi _T{=\tilde \beta 
\left( {1-x} \right)} / 2 )$.
 In the small doped case where $\left( {1-x} \right)>\!>{{t_0} / \tilde \beta}$ is satisfied, 
the large energy gap appears and 
the hybridizations between the bands are negligible. 
The lowest band is therefore constructed almost
only from the orbital described by Eq.(\ref{eqn : eq40}). 
 On the other hand, in sufficiently large doped case $\left( {1-x} \right)
\stackrel{\scriptstyle <}{\scriptstyle \sim} {{t_0}/
{\tilde \beta}}$, the energy gap shrinks and 
the orbital polarization is reduced. 
 This change of the band gap with varying the hole concentration 
is able to be detected as the inter-band transition in the optical measurements.
\par
Next, we demonstrate how the orbital structure controls the dimensionality 
of DOS. 
\begin{figure}[p]
\caption{}
\label{fig : DOS}
\end{figure}
\noindent
In Fig.\ref{fig : DOS}, 
we present DOS calculated in the several orbital structures. 
The spin structure is assumed as the F-type and the 
hole concentration is fixed as $x=0$. 
At first, we focus on the case for orbital 
$G\!\!:\!\!\left( {x^2-y^2} \right)/ \left( {3z^2-r^2} \right)$ and 
$F\!\!:\!\!\left( 
{x^2-y^2} \right)$ (Fig.\ref{fig : DOS} (b)) which are 
realized in $x<0.3$ in the ferromagnetic region as shown in Fig.\ref{fig : PD28}. 
Results in the both cases are essentially the same. 
DOS shows a two-dimensional character, because of 
absence of the hopping integral along the $c$-axis. 
In this case there is no difference in the kinetic energy between 
spin F and spin A, and $J_S$ favors spin A. 
 This causes a dip structure in the phase boundary 
between spin F and spin A at $x \sim 0.3$ and leads
an remarkable difference from the prediction by the conventional
double exchange model.
In Fig.\ref{fig : DOS} (c), 
DOS for the orbital $A\!\!:\!\!\,{{\left( {\left[ {3z^2-r^2} \right]
+\left[ {x^2-y^2} \right]} \right)} \mathord{\left/ {\vphantom {{\left( 
{\left[ {3z^2-r^2} \right]+\left[ {x^2-y^2} \right]} \right)} 
{\left( {\left[ {3z^2-r^2} \right]-\left[ {x^2-y^2} \right]} \right)}}} 
\right. \kern-\nulldelimiterspace} {\left( {\left[ {3z^2-r^2} \right]-
\left[ {x^2-y^2} \right]} \right)}}$ is shown. 
This orbital structure is stabilized in $0.3<x<0.8$ as shown in Fig.\ref{fig : PD28}. 
Although DOS should be essentially three dimensional one if there is 
no hybridization, results has two peaks 
which resembles that in the quasi one-dimensional system (Fig.\ref{fig : DOS}
(a)).
 This seems to be originated from the hybridization with the unoccupied band.
 For each case in Fig.\ref{fig : DOS}, the width in the lowest band is
the same, as expected in the case of 
$\left( {1-x} \right)>\!> {{t_0}/{\tilde \beta}}$.
 Therefore, by adjusting the orbital structure, 
the shape of DOS is modified and the center of mass for the occupied states
are changed so as to minimize the kinetic energy. 
From this viewpoint, a dimensionality of the lowest energy band 
plays an essential role on a gain of the kinetic energy. 
In the regions of $x \sim 1$, one-dimensional-like dispersion is advantageous 
as shown in Fig.\ref{fig : PD28}. 
\par
 Let us consider the case (B) 
in Fig.\ref{fig : PD1.21} where the more realistic energy parameters 
are adopted.
At the moment, a value of $J_S$ can not be estimated accurately, 
but there are two rough estimates.
 One is from the N\'eel temperature $T_N=130K$ for CaMnO$_3$ $(x=1.0)$, 
\cite{Woll}
which suggest $J_S = T_N/7.5 \cong  1.7 meV \cong 0.0023 t_0$
in the mean field approximation.
The fluctuations lower $T_N$, and hence increase the 
estimate for $J_S$. 
 Another estimate is obtained from the numerical calculations 
for LaMnO$_3$ $(x=0.0)$, which suggests
$J_S \cong 8 m eV \cong 0.011 t_0$. \cite{Ishi1}
 Although $J_S$ might depend on $x$ in real materials, 
we tentatively fix $J_S$ to be $0.009t_0$ represented by 
the broken line in Fig.\ref{fig : PD1.21}.
 Then the spin structure is changed as A $\to $ F $\to $ A$\to$ C $\to$ G, 
as $x$ increases, which is in good agreement with the experiments 
shown in Fig.\ref{fig : Experiment}(c).
\par
As we mentioned above, 
the ferromagnetic phase is roughly divided into the two regions: 
the low doped region ($x<0.3$) and high hole doped one ($x >0.3$). 
The former is not reproduced by the calculation without the 
Coulomb interaction between $e_g$ electrons, 
on other hand, the latter is not changed.
We conclude that the superexchange interaction discussed in the previous subsection
and the conventional double exchange interaction are 
dominant in the lower and higher doped regions, respectively. 
In the present formulation, 
character of the superexchange interaction in the metallic phase 
is derived by the following mechanism. 
We consider the state where the AF spin structure or the 
AF orbital structure exist. 
In this case, the eigen energy of $M_{kk';\gamma \gamma ';\alpha \beta }$ 
in Eq.(\ref{eqn : eq37})
corresponding to the Hartree-Fock energy 
is roughly expressed as 
%
$ E_k \sim \sqrt{\varepsilon(\vec k)^2+U_{e\!f\!f}(x)^2}  \ , $
%
where 
$\varepsilon(\vec k)$ and $U_{e\!f\!f}(x)$ 
are the diagonal and off-diagonal matrix elements of 
$M_{kk';\gamma \gamma ';\alpha \beta }$ , respectively. 
The former is the order of the transfer intensity 
and the latter is roughly estimated as 
$U_{e\!f\!f}(x)=U_c\varphi \sim U_c(1-x)$. 
$\varphi$ is the auxiliary field for the 
spin and/or iso-spin degrees and $U_c$ is the 
order of $U$.  
Therefore, the band width of the band is given by 
$w=E_{\vec k=\vec k_0}-U_{e\!f\!f}(x)$. 
$\vec k_0$ is the momentum where $\varepsilon(\vec k)=0$. 
It is approximated as $t^2/U_c$ for small $x$.  
This results is in contrast to the large $x$ case where 
the band width is the order of $t$. 
As de Gennes have pointed out 
in the case for small $x$, \cite{deGenne} 
the kinetic energy is determined by the product of the band width and 
the carrier concentration expressed in the present case as 
$ \Delta E \sim w x = (t^2/U_c) x \ .$
%
We stress that the energy scale $t^2/U_c$ 
corresponds to that in the superexchange interaction.
As increasing $x$, 
this ` superexchange character ' in the interaction is gradually 
replaced by the double exchange one.
Actually, the peak in the phase boundary at $x \sim 0.15 $ in 
Fig.\ref{fig : PD1.21} grows up with increasing $t/U_c$ (Fig.\ref{fig : A1} 
(a) ), 
on the other hand, the structure for $x>0.3$ is almost unchanged. 
\par
The modification of the F- and A-spin phases by changing the energy parameters discussed above 
explains the recent experiment 
in (La$_{1-z}$Nd$_z$)$_{1-x}$Sr$_x$MnO$_3$. \cite{Akimoto}
The phase transition between the spin A 
and spin F metallic-phases were studied by 
changing the band width, which is controlled by $z$, and the hole 
concentration $x$. 
With increasing the band width, 
the critical hole concentration $x_{F-A}$, where the phase transition occurs, 
is increased. 
\begin{figure}[p]
\caption{}
\label{fig : A1}
\end{figure}
\noindent
This experimental results is consistent with the present calculation  
shown in Fig.\ref{fig : A1} (b) where $t_0$ and $J_S \propto t_0^2$ 
are changed with fixing $\tilde \alpha$ and $\tilde \beta$. 
It is found that the critical carrier concentration $x_{F-A}$ is shifted to 
the higher $x$ region with increasing $t_0$ as consistent with the 
experiments. 
\par
 As shown in Fig.\ref{fig : PD28} and \ref{fig : PD1.21}, 
the orbital structure in the spin F phase is sensitively changed with the
energy parameters and the carrier concentration.
 This implies that there are many nearly degenerate orbital configurations 
in this phase. 
\begin{figure}[p]
\caption{}
\label{fig : Fluctuation}
\end{figure}
\noindent
In order to investigate the situation in detail, 
we compare the free energies with assuming several orbital structures in the 
ferromagnetic phase (Fig.\ref{fig : Fluctuation} (a)). 
We vary the angle in the orbital space $\theta$ with assuming the 
ferromagnetic orbital configuration. 
The calculation is also performed in the A-type AF spin case 
(Fig.\ref{fig : Fluctuation} (b)). 
 It is found that the energy variation is an order of magnitude smaller in the case of spin F
compared with the case of spin A.
 The difference between two cases are interpreted from the view point 
of the anisotropy of the electron transfer under the orbital ordering as follows. 
In spin A case, 
($x^2-y^2$) orbital is realized to maximize
the kinetic energy gain and the hopping along $c$-axis is forbidden. 
 In spin F case, on the other hand, such kind of lowering of 
the dimensionality does not occur because the three crystallographic axes are 
equivalent in this spin structure. 
As a result, the orbital configurations have more freedom. 
 This is the same physical idea behind the orbital liquid scenario proposed 
by present authors. \cite{Ishi2}
where the two-dimensional orbital fluctuation, 
characterized by the $\left( {x^2-y^2} \right)$, $\left( {y^2-z^2} 
\right)$ and $\left( {z^2-x^2} \right)$ orbital alignments, 
is suggested. 
 Through the orbital fluctuation among them, 
the kinetic energy in every direction is lowered and the ferromagnetic phase is stabilized. 
As a results, it is thought that the dip structure in the phase boundary between F- and A-spin structures 
shown in Fig.\ref{fig : PD1.21} 
disappear and the two ferromagnetic phases in the low and high doped regions are connected. 
\par
 Another possibility, which enhances the ferromagnetism 
in the region of $x<0.175$ is the JT distortion. \cite{Millis}
 In experiments, however, the static JT distortion disappears 
rapidly with increasing of $x$ around $x \sim 0.1$. \cite{Kawano1}
 Then the dynamical JT distortion should be considered in the metallic spin F state. 
According to the study in a large-$d$ model, \cite{Nagaosa}
the large Coulomb interaction is essential 
to explain both the isotope effect \cite{iso}
and the Raman scattering experiment. \cite{Kiichi}
 Then the dominant role of the Coulomb interaction assumed in this
paper seems to be reasonable.
\par
 One of the most remarkable results in Fig.\ref{fig : PD1.21} 
is an emergence of the spin A metallic phase for $0.2<x<0.5$ and the spin C 
phase for $x>0.5$. 
 These two phases are found in experiments using high quality
samples with narrower band width: Pr$_{1-x}$Sr$_x$MnO$_3$ and 
Nd$_{1-x}$Sr$_x$MnO$_3$. \cite{PrSr,Kuwahara1,Kawano2}
Also in La$_{1-x}$Sr$_x$MnO$_3$, spin A metallic is recently found
to emerge for $x>0.54$. \cite{mori}
It is worth noting that the metallic spin-A phase is highly contrast to the spin-A insulating 
phase observed around $x \sim 0$. 
An existence of the spin canting in the metallic spin-A 
phase is theoretically excluded as follows. 
 According to the study by de Gennes, \cite{deGenne} 
the spin canting from the insulating A-type spin structure 
lowers the energy by 
\begin{equation}
E_{cant}=E_{A}-\frac{Nb_{inter}^2}{8z\left|J_S\right|S^2}
{x^2} \ ,
\label{eqn : eq71}
\end{equation}
where $E_{cant}$ is the energy in the spin canted phase, 
$b_{inter}$ is the hopping integral along the $c$-axis, $z$ is the coordinate number 
along this axis, and $N$ is the number of the ion. 
The canting angle $\Theta$ is given by 
\begin{equation}
 \cos {{\Theta} \over 2}={{b_{inter}} \over {4\left| {J_S} 
 \right|S^2}}x \ .
\label{eqn : eq72}
\end{equation}
From the consideration, the spin A phase around $0<x<0.1$ in 
Fig.\ref{fig : PD1.21} is replaced by the spin canted phase
and a value of $J_S$ at the phase boundary is corrected downward as 
\begin{equation}
  J_S^{}\left( {F\!\!-\!\!Cant} \right)=J_S^{}\left( {F\!A} \right)
-\frac{Nb_{inter}^2}{36 z\left|J_S\right|S^2}
{x^2} \ .
\label{eqn : eq73}
\end{equation}
 On the other hand, in the spin A metallic phase
($0.2<x<0.45$) with the orbital structure $\left( {x^2-y^2} \right)$,
the hopping along the $c$-axis is forbidden by the orbital structure and 
hence no spin canting occurs.
 This is consistent with the recent neutron scattering experiment
showing no canting in this spin A metallic phase. \cite{Kawano2} 
Furthermore, 
the large anisotropy in the resistivity \cite{Kuwahara1} and 
the distortion of the $\rm MnO_6$ octahedron \cite{Kajimoto}
observed in the A-type AF at x=0.60 and 
the C-type AF at x=0.75 in Nd$_{1-x}$Sr$_x$MnO$_3$ 
are consistent with the calculated orbital structure: 
$\left( {x^2-y^2} \right)$ for spin A and  
$\left( {3z^2-r^2} \right)$ in spin C. 
%
%
%
%
\section{Discussion and Conclusions}
%
%
%
 Here we discuss the relation between the present work and 
the previous ones performed at $x=0.0$.
 Kugel and Khomskii \cite{Kugel} and Koshibae $et\ al$. \cite{Koshi} 
have dealt with the spin and the orbital orderings at $x=0$, 
taking into account the orbital degeneracy and 
the electron-electron interactions. 
 They used the effective Hamiltonian obtained by the 
second-order perturbative expansion in the limit of strong 
Coulomb repulsion.
 Kugel and Khomskii studied the 
ground state spin and orbital structures 
in the small limit of $J/U'$ without the 
antiferromagnetic interaction $J_{S}$ between $t_{2g}$ spins. 
In the mean field calculation, 
the A-type spin structure is reproduced but 
the orbital structure is almost 
$(z^2-x^2)/(y^2-z^2)$ type ordering, 
which disagrees with one expected from a type 
of the observed JT distortion. 
On the other hand, Koshibae $et\ al$. also studied the spin and orbital 
structures in the large limit of $J/U'$ with taking into 
account $J_S$. 
By using the exact diagonalization method in the finite 
cluster system, 
it is found that the spin correlation changes
as $F \rightarrow A \rightarrow C \rightarrow G$-type
as $J_S$ increases.
They also found that in spin A phase, 
component of ${{\left( {3x^2-r^2} \right)} \mathord
{\left/ {\vphantom {{\left( {3x^2-r^2} \right)} {\left( {3y^2-r^2} 
\right)}}} \right. \kern-\nulldelimiterspace} {\left( {3y^2-r^2} 
\right)}}$ 
or 
${{\left( {z^2-x^2} \right)} \mathord
{\left/ {\vphantom {{\left( {z^2-x^2} \right)} {\left( {y^2-z^2} 
\right)}}} \right. \kern-\nulldelimiterspace} {\left( {y^2-z^2} 
\right)}}$ orbital alignments are enhanced,
although both components are not distinguished 
in the orbital correlation function calculated there. 
 Our calculation covers both of these two cases with
the following features;
a) $J_{S}$ is taken into account.
b) perturbative expansion is not used, $i.e.$, applicable
for any parameters, $\tilde \alpha / \tilde \beta$,
${J \mathord{\left/ {\vphantom {J {U'}}} \right. \kern-
\nulldelimiterspace} {U'}}$.
c) spin and the orbital orderings in the infinite system are studied. 
As shown in Figs.1-3, 
in the limit of $\tilde \alpha / \tilde \beta >\!>1$, 
a sequential change of the spin structure with changing 
$J_S$ is observed. 
The orbital structure is also rearranged and 
only in the spin A-AF phase, 
$(\theta_I,\theta_{I\!I})=(90^\circ,270^\circ)$ 
is stabilized. 
On the hand, in the limit of $\tilde \alpha / \tilde \beta  \sim 1$, 
$(\theta_I,\theta_{I\!I})=(120^\circ,-120^\circ)$ 
is stabilized in the spin A-AF phase. 
Therefore, 
the results obtained in the previous calculation \cite{Kugel,Koshi}
are reproduced by the present calculation in the unified fashion. 
In addition, our calculation shows that the orbital 
ordering expected from a type of JT distortion can not be realized for 
any value of $\tilde\alpha/\tilde\beta$.
It implies indispensability of
the anharmonicity from the JT distortion \cite{Kanamori}
for proper description at $x=0$. 
 This is in accordance with the results by 
the Hartree-Fock \cite{Mizo} 
and the first-principle calculation. \cite{Hamada}
except that spin A phase 
is realized even without JT shown as in Fig.\ref{fig : PD1.21}.
%
%
\par
Next, let us pay our attention to the doped case. 
As mentioned above, 
the perfectly polarization of the orbital moment 
derived by the electron-electron interactions 
plays an essential role on the spin ordering. 
It controls the dinemsionality of the 
conduction band through the anisotropy 
of the transfer intensity. 
If not, the anisotropy of the conduction band 
is weakened, because of the large hybridyzation 
between the lowest band and the other ones. 
\begin{figure}[p]
\caption{}
\label{fig : opara}
\end{figure}
\noindent
 As a comparison we show in Fig.\ref{fig : opara} a phase
diagram without the orbital polarization by assuming $\beta=0$.
 The phase diagram is dominated by the ferromagnetic state for
reasonable values of $J_S$, and the nonmonotonic behaviors 
shown in Figs.\ref{fig : PD28} and \ref{fig : PD1.21} disappear.
 We conclude that the almost saturated orbital polarization
is essential to obtain the experimentally observed phase diagram
and the unique character observed in the each spin phase, 
$e.g.$, the two dimensional conduction and the no spin canting in the metallic A-AF phase. 
 For such a large orbital polarization in the metallic phase, 
the strong Coulomb interaction 
is indispensable rather than the JT coupling.
%

 In summary, we have studied the phase diagram of R$_{1-x}$A$_x$MnO$_3$ 
at zero temperature in the plane of $x$ 
(hole concentration) and $J_S$(AF exchange interaction between 
$t_{2g}$ spins) in the mean field approximation. 
 The global features are understood in terms of the superexchange 
interaction and the double exchange interaction, 
which is considerably modified from the conventional one 
due to the strong correlation and 
the orbital degeneracy. 
The large orbital polarization 
originated from the electron-electron interaction 
is indispensable to reproduce the phase diagram 
experimentally observed. 
\par
 The authors would like to thank S. Maekawa, Y. Tokura, K. Terakura, 
Y. Moritomo, and I. Solovyev for their valuable discussions. 
 This work was supported by Priority Areas Grants from the Ministry 
of Education, Science and Culture of Japan, and the New Energy and 
Industrial Technology Development Organization (NEDO).


\par
\noindent
$^{*}$  Present address: 
Institute for Materials Research, Tohoku University, 
Sendai, 980-8577, JAPAN. 
%
%

%
%
\vfill 
\eject
\noindent
Figure captions
\par
\noindent
\\
Figure \ref{fig : Experiment}. 
 Phase diagrams of (a) La$_{1-x}$Sr$_x$MnO$_3$,\cite{LaSr1} 
(b) Pr$_{1-x}$Sr$_x$MnO$_3$, \cite{PrSr} and 
(c) Nd$_{1-x}$Sr$_x$MnO$_3$. \cite{Kuwahara1}
\par
\noindent
\\
Figure \ref{fig : Free}.
 Free energies for each spin alignment 
as a function of the antiferromagnetic interaction between
$t_{2g}$ spins $J_S$ at $x=0$. 
The energy parameters are chosen to be $\tilde \alpha=70$ 
and $\tilde \beta =2.5$ (case (A)). 
Orbital structures are also noted.
\par
\noindent
\\
Figure \ref{fig : SOdep}.
 Orbital structures at $x=0$ as a function 
of the antiferromagnetic interaction between the $t_{2g}$ 
spins $J_S$. 
The energy parameters are chosen to be $\tilde \alpha=70$ 
and $\tilde \beta =2.5$ (case (A)). 
\par
\noindent
\\
Figure \ref{fig : thetadep}.
Orbital structures in each sublattice 
at $x=0$ 
as a function of the electron-electron interaction parameters 
$\tilde \alpha/\tilde \beta$. 
The A-type spin alignment and the G-type orbital one are assumed. 
\par
\noindent
\\
Figure \ref{fig : OrbforA}.
 Orbital orderings described as $\left( {\theta _I,
\theta _{I\!I}} \right)$=($120^\circ,-120^\circ$) and
($60^\circ,-60^\circ$) 
respectively. 
\par
\noindent
\\
Figure \ref{fig : JT-1}.
 Orbital structures in the each sublattice as a  
function of magnitude of the JT splitting ${{gr}/{t_0}}$.
The energy parameters are chosen to be $\tilde \alpha=70$ 
and $\tilde \beta =2.5$ (case (A)). 
$r$ is fixed to be 0.028. 
\par
\noindent
\\
Figure \ref{fig : JT-2}.
 Variation of the phase boundary between spin F- and spin A-phases
as a function of (a) the magnitude of the JT distortion $r$  
with fixing the JT splitting to be $g=0$ (left-hand side panel), and 
(b) the JT splitting ${{gr}/{t_0}}$ with fixing the magnitude of the 
distortion to be $r=0.028$ (right-hand side panel).
\par
\noindent
\\
Figure \ref{fig : JT-4}.
 Magnitude of the transfer integrals for each orbital 
orderings, appearing in the calculation of Eq.(\ref{eqn : eq68}).
\par
\noindent
\\
Figure \ref{fig : PD28}.
 Mean field phase diagram as a function of the 
carrier concentration ($x$) and the antiferromagnetic interaction 
between $t_{2g}$ spins ($J_S$ ). 
The energy parameters are chosen to be $\tilde \alpha=70$ 
and $\tilde \beta =2.5$ (case (A)). 
Schematic orbital structure in the each phase 
is also shown. 
\par
\noindent
\\
Figure \ref{fig : PD1.21}.
 Mean field phase diagram as a function of the carrier 
concentration ($x$) and the antiferromagnetic interaction 
between $t_{2g}$ spins ($J_S$). 
The energy parameters are chosen to be $\tilde \alpha=8.1$ 
and $\tilde \beta =6.67$ (case (B)). 
Schematic orbital structure in the each phase 
is also shown. 
Dotted line ($J_S$ =0.009) 
well reproduces the change of the spin structure 
experimentally observed (see text).
\par
\noindent
\\
Figure \ref{fig : band-str}.
 Splitting of the band structure for $e_g$ electrons.
The up and down arrows represent the direction of spin, while
$\left| a \right\rangle$ and $\left| b \right\rangle$ represent 
one of the $e_g$ orbitals in each band. (a) $\tilde\alpha=\tilde\beta
=0$ case, (b) $\tilde\alpha \ne 0$, $\tilde\beta=0$ case, and (c)
$\tilde\alpha \ne 0$, $\tilde\beta \ne 0$ case.
\par
\noindent
\\
Figure \ref{fig : DOS}.
Density of states of the lowest band at $x=0$. 
The orbital structures are assumed to be 
(a) $\theta= 180^\circ$ ($d_{3z^2-r^2}$) , 
(b) $\theta= 0^\circ$ ($d_{x^2-y^2}$), 
and (c) $\theta= 90^\circ$. 
The ferromagnetic spin structure is assumed. 
The energy parameters are chosen to be $\tilde \alpha=70$ 
and $\tilde \beta =2.5$ (case (A)).
A Value beside each bond represents
the transfer intensity.
\par
\noindent
\\
Figure \ref{fig : A1}.
 Variation of the phase boundary $J_S(FA)$ 
between the spin F and A-phases.  
(a) Values of the parameters $\tilde\alpha$ and $\tilde\beta$ are changed.  
(b) Values of the parameter $t_0$ is changed. 
The cross point $x_{FA}$ represents the transition from Spin F 
to Spin A, where $J_S$ scales as $J_S \propto t_0^2$
\par
\noindent
\\
Figure \ref{fig : Fluctuation}.
The energy as a function of the orbital state characterized by $\theta$ in 
the several value of $x$. (a) Spin F is assumed. (b) Spin A is assumed. 
In both cases, the orbital F-type structure is assumed. 
The energy parameters are chosen to be $\tilde \alpha=8.1$ 
and $\tilde \beta =6.67$ (case (B)).
\par
\noindent
\\
Figure \ref{fig : opara}.
 Mean field phase diagram calculated with 
assuming $\tilde\beta=0$.
\par
\noindent
\par 
\noindent
%
%
\end{document}